\documentclass[useAMS,usenatbib,]{mn2e}
\usepackage{graphicx}
\usepackage{subfigure}

\def \NII {[N~{\sc ii}]6584~\AA}
\def \ha {H$\alpha$}
\def \vhel{\ifmmode{~V_{{\rm HEL}}}\else{~$V_{{\rm HEL}}$}\fi}
\def\msun{\ifmmode{{\rm\ M}_\odot}\else{${\rm\ M}_\odot$}\fi}

\def\myr{\ifmmode{{\rm\ M}_\odot{\rm\ yr}^{-1}}
         \else{\mbox{${\rm\ M}_\odot$ yr$^{-1}$}}\fi}

\def\tena#1 #2 {\ifmmode{#1 \times 10^{#2}}\else{\mbox{$#1 \times 10^{#2}$}}\fi}
\def\kms{\ifmmode{~{\rm km\,s}^{-1}}\else{\mbox{~km s$^{-1}$}}\fi}

\title[Kinematics of the ring-like nebula SuWt 2]{Kinematics of the ring-like nebula SuWt 2\thanks{Based on observations made with ESO telescopes at the La Silla or Paranal Observatories under programme IDs 74.D-0373 \& 55.D-0550.}}
\author[D. Jones et. al]{D. Jones,$^{1,2}$\thanks{E-mail:
david.jones-3@postgrad.manchester.ac.uk} M. Lloyd,$^{1}$ D.L. Mitchell,$^{1}$ D.L. Pollacco,$^{3}$ T.J. O'Brien,$^{1}$ 
\newauthor
and N.M.H. Vaytet.$^{4}$
\\
$^{1}$Jodrell Bank Centre for Astrophysics, University of Manchester, M13 9PL, UK\\
$^{2}$Isaac Newton Group of Telescopes, Apartado de Correos 368, E-38700 Santa Cruz de La Palma, Spain\\
$^{3}$Astrophysics Research Centre, Queen's University Belfast, BT7 1NN, UK\\
$^{4}$Service d'Astrophysique, CEA/DSM/IRFU/SAp, Centre d'\'{E}tudes de Saclay, L'Orme des Merisiers, 91191 Gif-sur-Yvette Cedex, France}

\begin{document}

\date{Accepted xxxx xxxxxxxx xx. Received xxxx xxxxxxxx xx; in original form xxxx xxxxxxxx xx}

\pagerange{\pageref{firstpage}--\pageref{lastpage}} \pubyear{2008}

\maketitle

\label{firstpage}

\begin{abstract}
We present the first detailed spatio-kinematical analysis and modelling of the Southern planetary nebula SuWt 2.  This object presents a problem for current theories of planetary nebula formation and evolution, as it is not known to contain a central post-main sequence star.

Deep narrowband \NII\ images reveal the presence of faint bipolar lobes emanating from
the edges of the nebular ring.  Longslit observations of the \ha\ and \NII\ emission lines were obtained using EMMI
on the 3.6-m ESO-NTT.  The spectra reveal the nebular morphology as
a bright torus encircling the waist of an extended bipolar structure.  By deprojection, the inclination of the
ring is found to be 68$\degr$ $\pm$ 2$\degr$ (c.f. $\sim$90$\degr$ for
the double A-type binary believed to lie at the centre of the nebula),
and the ring expansion velocity is found to be 28 \kms{}.

Our findings are discussed with relation to possible formation
scenarios for SuWt~2.  Through comparison of the nebular heliocentric
systemic velocity, found here to be $-25 \pm 5 \kms$,  and the heliocentric systemic velocity of the double A-type binary, we conclude that neither component of the binary could have been the nebular progenitor.  However, we are unable to rule out the presence of a third component to the system, which would have been the nebula progenitor.
  
\end{abstract}


\section{Introduction}
Planetary nebulae (PNe) represent a late stage in the evolution of intermediate mass stars, the production of which has long been thought to be the result of a dense stellar wind, originating from the progenitor Asymptotic Giant Branch (AGB) star, being swept up into a fine rim by a faster wind from the emerging White Dwarf (WD) \citep{kwok78}. This model was later extended to become the Generalised Interacting Stellar Winds (GISW) theory, stating that enhanced equatorial mass-loss in the slow, dense wind phase can lead to a bipolar shape in the resulting nebula (see e.g., review by \citealp{balick02}).  The mechanism behind this highly non-isotropic mass-loss represents a particularly hot topic in the field of PN research.  One favourable mechanism is for the central star of the planetary nebula (CSPN) to undergo a common envelope (CE) evolution with a binary partner.  For a CE to form, one component of a close-binary system must overflow its Roche lobe and begin to accrete on to its partner, and the timescale for mass transfer must be considerably shorter than the timescale on which the accretor can thermally adjust, thus causing the accretor to also fill its Roche lobe.  Any further mass lost by the donor, will then go on to form a CE which surrounds both stars.  The shedding of this CE, by angular-momentum transfer, $\alpha$-$\omega$ dynamo or accretion-driven jets \citep{nordhaus06}, provides the mass anisotropy required to form bipolar nebulae.  The hydrodynamic simulations of \citet{rasio96} confirm that equatorial density enhancement is a natural consequence of a CE evolution.

\begin{figure*}
\centering
\includegraphics{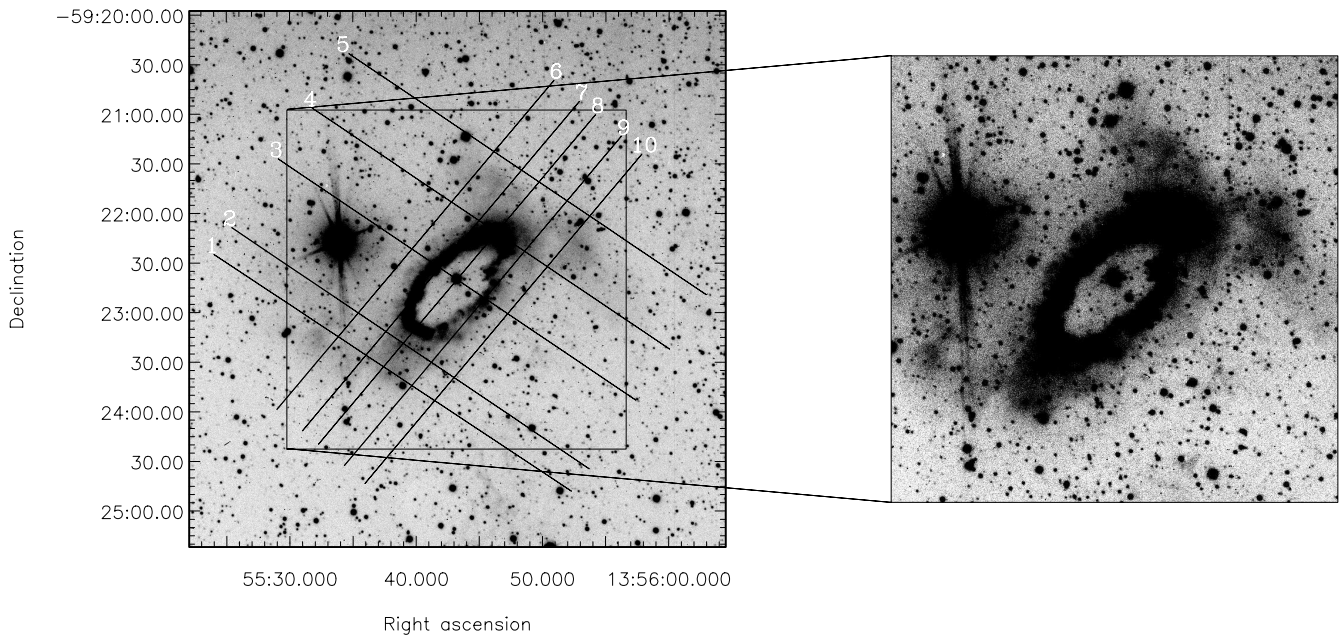}
\caption{ESO-NTT exposure of SuWt 2 in the light of \NII\ showing the position and extent of slits used, the cut-out shows the same image at greater contrast in order to highlight the extent of the nebular ring and the presence of the bipolar lobes.}
\label{fig:slits}
\end{figure*}

SuWt~2 (PN~G311.0+02.4, \mbox{$\alpha = 13^h55^m43.23^s$}, \mbox{$\delta = -59\degr22'40.03''$} J2000) is described by \citet{schuster76} as `an elliptical, nebular ring upon which several starlike images are superposed'.  \citet{west76} classified SuWt~2 as a PN based on of the presence of strong emission from forbidden lines typical of this class of object.  \cite{smith07} added further evidence through the comparison of line strengths between \NII\ and \ha{}, concluding that the nebula is nitrogen-rich and as such originates from a post-main sequence object.  They also derive an electron temperature from the [N~{\sc ii}] line ratio of T$_e=11,400$~K, which is similarly consistent with those of other planetary nebulae \citep{kaler86}.

Photometric analysis of the CSPN reveals a 4.9 day period, eclipsing
binary \citep{bond02}.  Later spectroscopic analysis confirmed this
period and also revealed the two stars to both be A-type with masses
of about 3 M$_\odot$; there is no indication of a hotter component in the
observed spectra \citep{exter03b}.  It is unclear what mechanism could
lead to the formation of a planetary nebula associated with this
system.  It is also impossible for either star to account for the
ionising flux required to illuminate the nebula, leading
\citet{bond02} to put forward the possibility that the system is
actually a triple with a distant, and as yet undetected,  third WD
component which would account for both the origin and illumination of the nebular shell.  However, \citet{smith07} speculate  that the source of the ionising radiation could be the bright B2 star approximately 1\arcmin{} Northeast of SuWt~2.  This hypothesis could also account for the observed brightness enhancement in the Northeastern edge of the nebular ring and would require the star to be at a similar distance to SuWt~2. They conclude that further investigation is required to assess the validity of this hypothesis.  It is also possible that the nebula is actually seen in recombination \citep{exter03b}, meaning no current source of ionising radiation is required.  

In this paper we present longslit spectroscopy of SuWt~2, from which, combined with deep narrow-band images (see Figure \ref{fig:slits}), we derive a spatio-kinematical model of the nebula, with the aim of probing the nature of the relationship between the nebula and the double A-type system apparently at its centre.

\begin{figure*}
\includegraphics{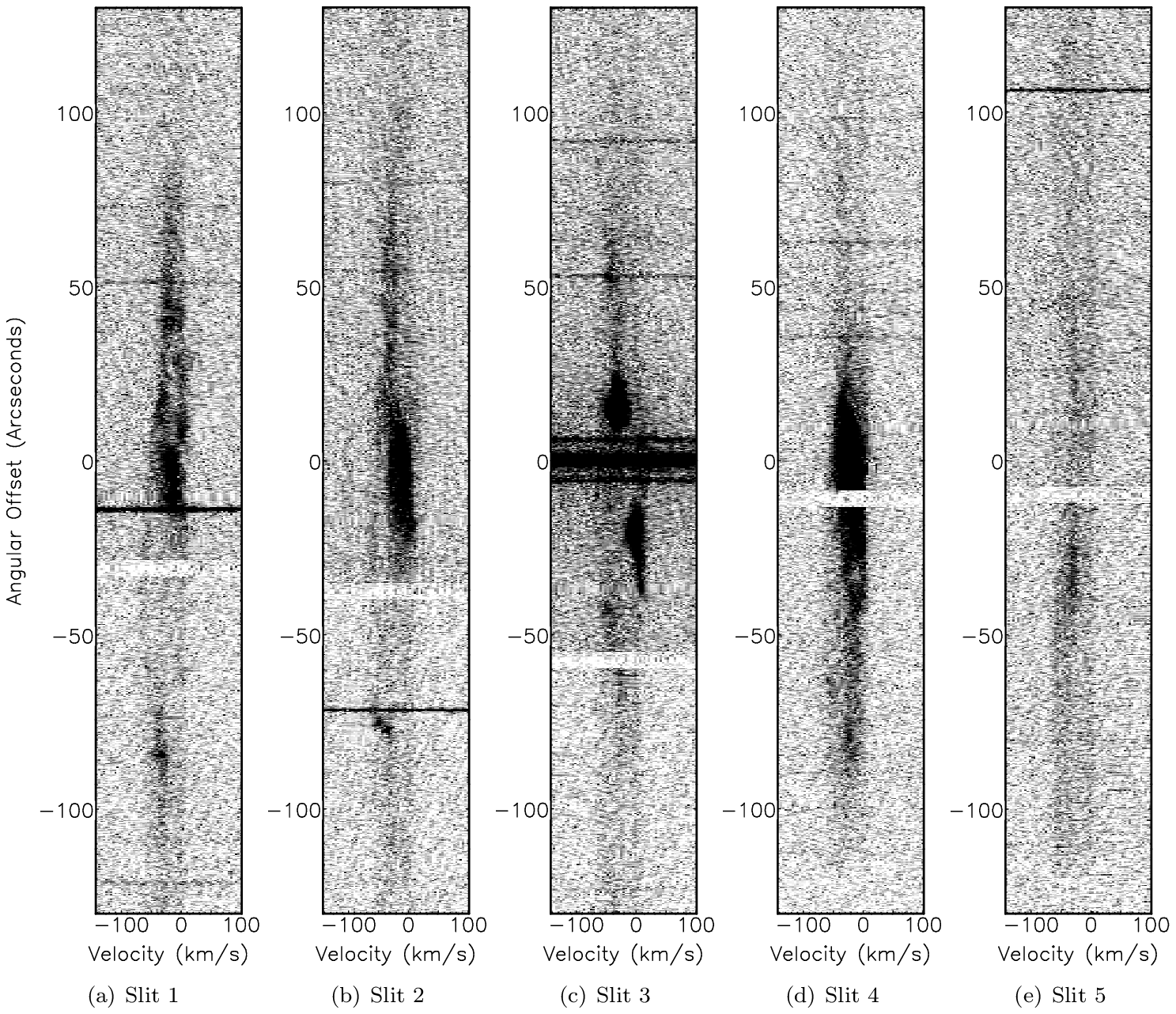}
\caption{The observed \NII\ PV arrays from slits 1 to 5 (see Figure
  \protect\ref{fig:slits}).   Note that the gap between the two CCD chips appears as a white band at negative angular offsets.}
\label{fig:slits1-5}
\end{figure*}

\section{Observations and Data Reduction}
\label{sec:obs}

The narrow band \NII\ image of SuWt~2 shown in Figure \ref{fig:slits} was obtained using the ESO Multi-Mode Instrument (EMMI, \citealt{dekker86}) on the 3.6-m ESO New Technology Telescope (NTT) on 
1995 April 20 with an exposure time of 1000~s and seeing of 1\arcsec{}
(the pixel scale of EMMI in this mode is 0.27\arcsec{} per pixel).
The image shows the bright, nebular ring and much fainter lobes
extending to the Northeast and Southwest of the ring, which were first
alluded to by \citet{exter03b} but have never before been presented in
such a deep image.  The ring is almost elliptical, but somewhat
irregular and slightly wider to the Northwest.  It can be seen, by
comparing the two panels in Figure \ref{fig:slits}, that the bright
nebular ring appears much thicker and more regular when shown at high
contrast, as noted by \citet{smith07}.  The lobes seen to be
protruding to the Northeast and Southwest of the ring appear to show
the bipolar morphology typical of many ring-PNe (for
example; IC 2149: \citealp{vazquez02}, Me 1-1: \citealp{pereira08} \&
WeBo 1: \citealp{bond03}).  The nebula can be seen to exhibit
brightness variations across not only the ring, as noted by
\cite{smith07}, but also in the bipolar lobes; both lobes appear
brighter along their Northwestern edges and the lobe protruding to
the Northeast appears brighter than its Southwestern partner. \cite{west76} suggested that the observed Balmer decrement indicated some
obscuration, particularly in the Western part of the ring.
Furthermore, the bright star appearing inside the nebular ring, the
double A-type binary, can clearly be seen to be offset by
approximately 4\arcsec\ North and 2\arcsec\ East from the ring's centre as found by ellipse fitting.

Spatially resolved, longslit emission-line spectra of SuWt~2 have been obtained with EMMI on the NTT. Observations took place in 2005 March 2-4 using the red arm of the spectrograph which employs two MIT/LL CCDs, each of 2048 $\times$ 4096 15 $\mu$m pixels ($\equiv$ 0.166\arcsec\ per pixel), in a mosaic. There is a gap of 47 pixels ($\equiv 7.82\arcsec$ ) between the two CCD chips which can be seen in the observed spectra.
 
EMMI was used in single order echelle mode, with grating \#10 and the narrowband \ha\ filter (\#596) to isolate the $87^{\rm{th}}$ echelle order containing the \ha\ and \NII\ emission lines.  Binning of \mbox{2 $\times$ 2} was used giving spatial and spectral scales of 0.33\arcsec\ per pixel and 3.8\kms\ per pixel, respectively.  The slit had a length of 330\arcsec{} and width 1\arcsec{} (\mbox{$\equiv$ 10 km~s$^{\rm -1}$}).  All integrations were of 1800~s duration and the seeing never exceeded 1\arcsec{}.

Data reduction was performed using \textsc{starlink} software. The spectra were bias-corrected and cleaned of cosmic rays. The spectra were then wavelength calibrated against a long exposure ThAr emission lamp, taken at the start of each night.  The calibration was confirmed using short Ne emission lamp exposures throughout the night, and by checking the wavelengths of skylines visible in the exposure. Finally the data were rescaled to a linear velocity scale (relative to the rest wavelength of \NII\ taken to be $6583.45$~\AA{}) and corrected for Heliocentric velocity.

\begin{figure*}
\includegraphics{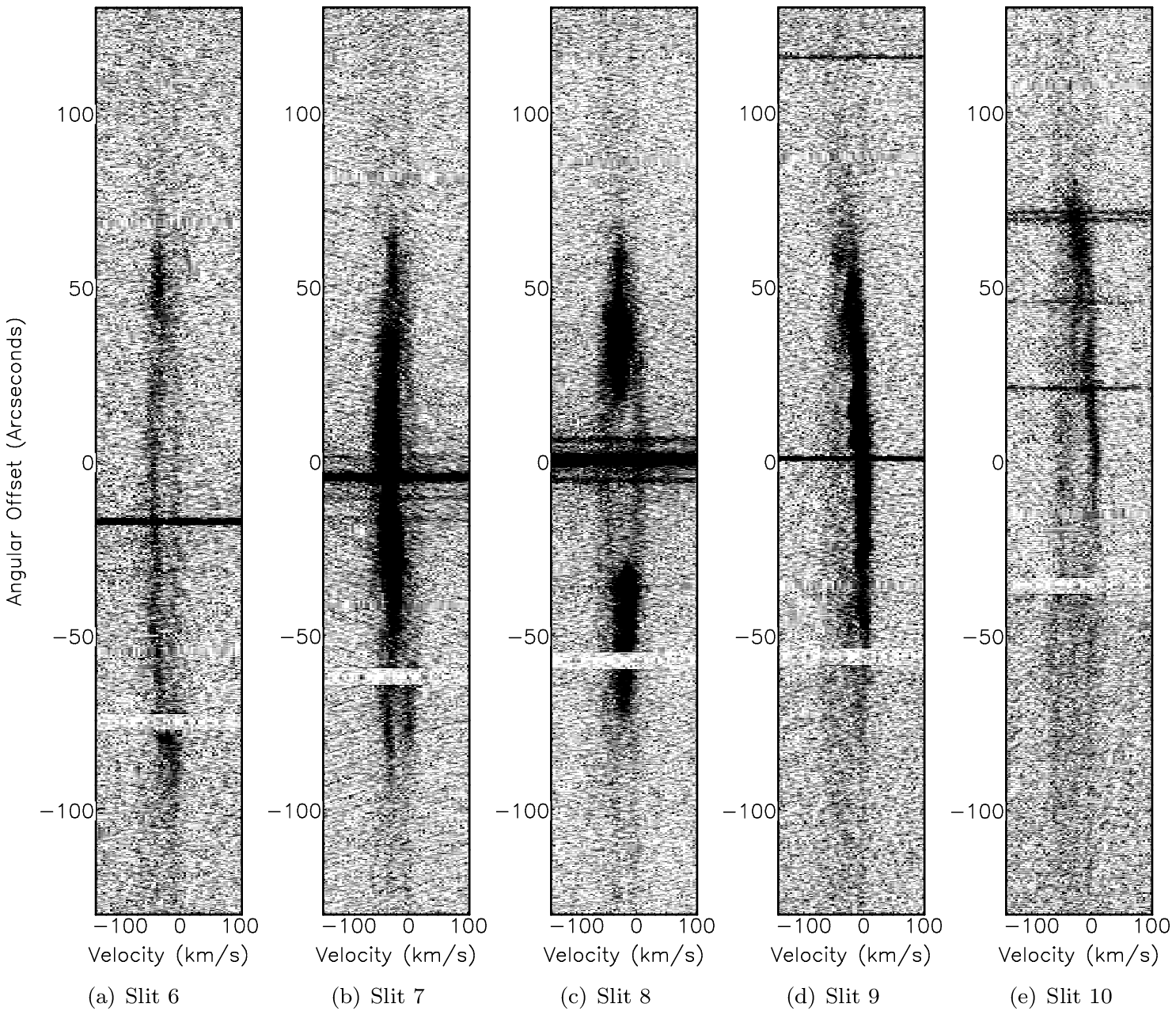}
\caption{The observed \NII\ PV arrays from slits 6 to 10 (see Figure
  \protect\ref{fig:slits}).   Note that the gap between the two CCD chips appears as a white band at negative angular offsets.}
\label{fig:slits6-10}
\end{figure*}

 In total, ten integrations were obtained, five inclined at 9$\degr$ to the minor axis of the nebular ring at a \mbox{PA = 56$\degr$} (numbered 1 to 5)  and five inclined at 87$\degr$ to the minor axis of the nebular ring at a \mbox{PA = $-$40$\degr$} (numbered 6 to 10). The slit positions are shown on the deep image of SuWt~2 in Figure \ref{fig:slits}. The fully reduced position-velocity (PV) arrays for  \NII\ emission are shown in Figures \ref{fig:slits1-5} and \ref{fig:slits6-10}.

 The y-axis zeroes (spatial dimension) of the PV arrays are set to the
 point at which the slit crosses either the major (slits 1 to 5) or
 the minor (slits 6 to 10) axis of the nebular ring.  Here both the
 major and minor axes are taken to be through the bright double A-type
 binary star.
Note that the spectrum of the double A-type binary is not spatially centred 
 between the bright regions of emission from either side of the
 nebular ring [Figure \ref{fig:slits6-10}(c)], confirming that the
 binary does not lie at the geometric centre of the nebula.
  Although slits 7 and 9 [Figures \ref{fig:slits6-10}(b) and (d)] show
 stellar continuum at approximately 0\arcsec{}, these are merely
 coincidental field stars and only slits 3 and 8 [Figures \ref{fig:slits1-5}(c) and \ref{fig:slits6-10}(c)] actually cross the double A-type binary.  Similarly both slits 3 and 8 appear to show three stellar continua around zero arcseconds, this is actually an artefact of the system resulting from the comparative brightness of the double A-type binary relative to the nebular emission.

It can be seen from the PV array of slit~3 [Figure \ref{fig:slits1-5}(c)] that the nebular ring is
 tilted such that its Northeastern side is towards the observer
 (assuming that it is expanding radially outwards). The PV array from
 slit~8 [Figure \ref{fig:slits6-10}(c) appears to show a `velocity
 ellipse' with very bright end
components where the slit crosses the bright ring, and much fainter
 red and blue components connecting the bright ends. 
The bright emission in slits 7 and 9 [Figures \ref{fig:slits6-10}(b)
 and (d)], which cross the Northeast and Southwest limbs of the ring, 
 again show that the Northeastern
part of the ring is blue-shifted and the Southwestern part is
red-shifted.

The PV array from slit~1 [Figure \ref{fig:slits1-5}(a)] shows that emission from both the brighter
 Northeastern lobe and the fainter Southwestern lobe is split into two components out to at least
 50\arcsec. This splitting can also be seen in slits 2 to 5, [Figure \ref{fig:slits1-5}],
 particularly in the Southwestern lobe. Slits 6 and 10 [Figure
 \ref{fig:slits6-10}(a) and (e)], which do not cross the bright ring,
 show velocity ellipses in their PV arrays, indicating that the two
 lobes have a hollow-shell structure.

Other noticable features include a compact
 bright component in slits 1 at $-$80\arcsec\ and 2 at
 $-$70\arcsec\ (Figure \ref{fig:slits1-5}(a) and (b)), and a double
 component in the emission from the region just beyond the
Southeastern edge of the ring which is most apparent in 
 Slit 7 [Figure \ref{fig:slits6-10}(b)].
Faint emission  can be seen extending out beyond $\pm 100$\arcsec\
 along all the slits.

\begin{figure*}
\includegraphics{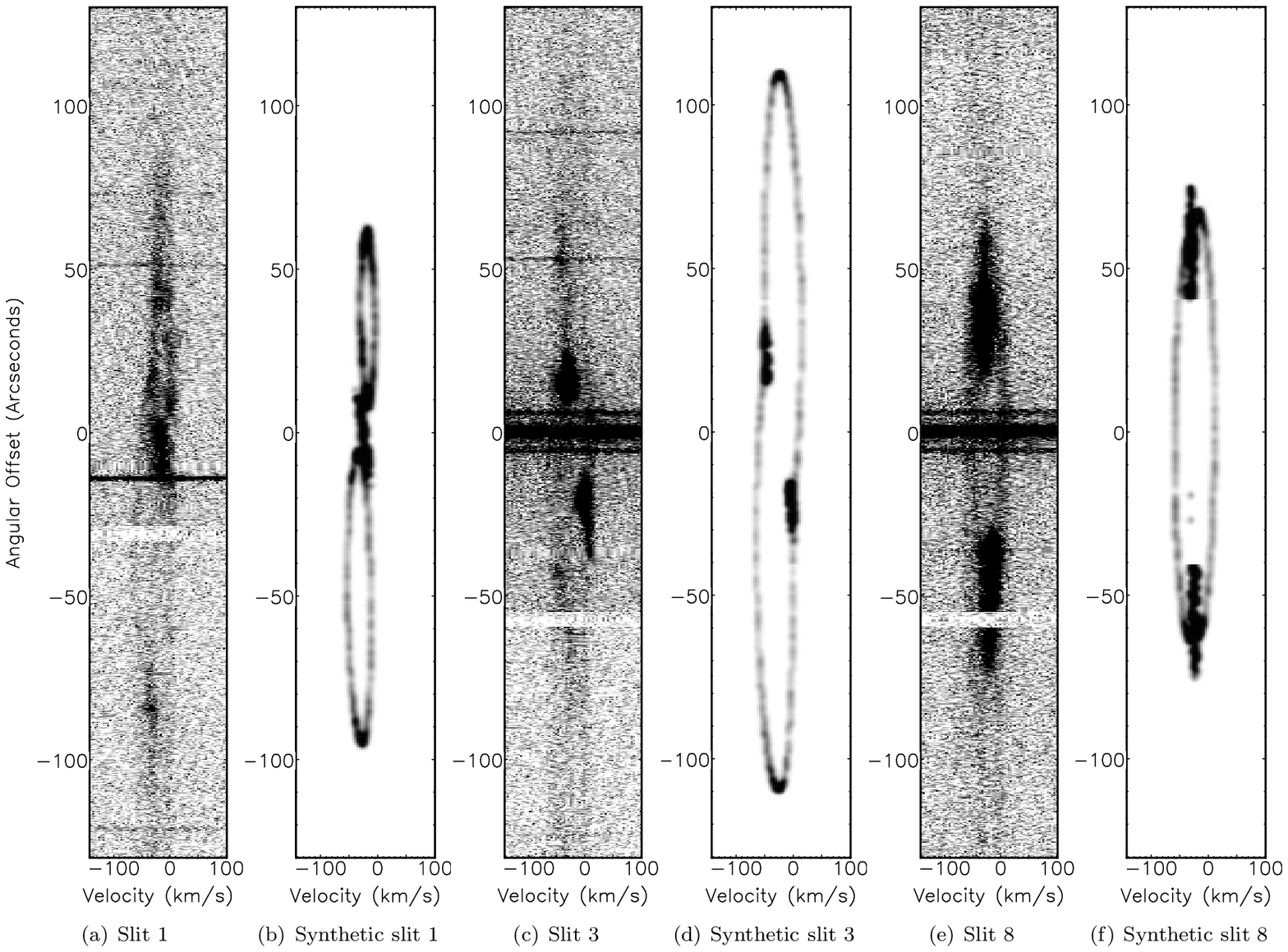}
\caption{A selection of the synthetic PV
  arrays derived from the closed bipolar model, shown with the
  corresponding observed \NII\ arrays.}
\label{fig:modelcomp}
\end{figure*}

\section{Analysis}
\label{sec:analysis}

A spatio-kinematic model of the \NII\ emission from SuWt~2 has been
  derived in order to confirm that its structure is indeed bipolar and
  to constrain the derived systemic velocity so that this can be
  compared to the systemic velocity of the double A-type binary.

The model was developed using \textsc{shape} \citep{steffen06} and compared to the observed \NII\ emission, which
has a significantly greater intensity and less thermal broadening than
the \ha\ emission.  The observed \ha\ profiles are also contaminated by background galactic emission. 

As a starting point, it was assumed that the visible
ring of SuWt~2 is, in fact, circular, and by deprojection an
inclination of \mbox{68$\degr$ $\pm$ 2$\degr$} was derived.  This
value is in reasonable agreement with the value of \mbox{64$\degr$
  $\pm$ 2$\degr$} found by \citet{smith07} who used the same method
but a different image.  The model nebula consists of a
bright torus (giving rise to the observed ring) and fainter, symmetric, bipolar
lobes. A Hubble-type flow was assumed, with the same scale velocity for each nebular component (torus and bipolar lobes).
 The model parameters (size, shape, expansion scale velocity, etc.)
 were manually varied, and the results compared by eye to the
 observations, both spectral and imaging, until a best-fit was
 found.
 The best-fit model has a scale velocity 
 equivalent to a 28\kms\ expansion velocity (from the nebular centre)
 for the torus and a systemic velocity of $-25 \pm 5\kms$. 
Both open and closed bipolar lobe models were tested, the observed
spectra from the lobes are too faint and irregular to conclusively distinguish between these two alternatives.  It seems that the Northeastern
lobe is better matched by a closed model whereas the Southwestern
lobe is better matched by an open model. It is not unheard of to find nebulae with components inclined with respect to each other (for example; Menzel 3 - \citealp{santander04}), however there is no evidence that the
lobes are inclined at a significantly different angle to the ring.

A selection of synthetic PV arrays extracted from the closed lobe
model are shown in Figure
\ref{fig:modelcomp}, together with the corresponding observed data. 
The synthetic arrays have been
convolved in both spatial and spectral dimensions to reflect the
observed spatial and spectral resolutions.

\subsection{Comparison of model to data}
The bright emission from the ring is generally well reproduced by the
synthetic data (Figure \ref{fig:modelcomp}), although
the model torus appears at first sight to be larger than the observed
bright ring.  This is because the model torus has dimensions matching
the full extent of the ring, but does not include the significant
brightness variations observed across the ring.  The spectra confirm
that the emission from the faint, outer part of the ring is
kinematically consistent with the emission from the bright, inner
part.
 The model PV array from slit 8, [Figure \ref{fig:modelcomp}(f)],
 which lies close to the major axis of the ring,
 reveals that the bright, end components of the `velocity ellipse' are from the bright ring, whereas the connecting faint
components are in fact from the much fainter bipolar lobes.
 The faint emission from the lobes can also be 
seen on slits 7 and 9 [figure \ref{fig:slits6-10}(b) and (d)], where slit 7 shows the
blue-shifted edge of the bright ring, and a faint red-shifted
component from the red-shifted lobe behind it, and slit 9 shows the
red-shifted edge of the bright ring and the faint blue component from
the near-side lobe which lies in front of it.  The identification of
the observed ring as a flattened equatorial ring structure rather than
a limb-brightened ellipsoid is consistent with the findings of
\cite{smith07}, based on their intensity profile analysis of the image.

  The bright emission around 0\arcsec\ offset from slit 1 [Figure \ref{fig:modelcomp}(a)] is seen to
  originate in the outer edge of the torus and would not be matched by
  a model with a smaller torus.  On the irregular Northwestern part of the ring, some emission appears from within the region that would be occupied by the model torus.  However if the torus were to be offset to the Southeast to accommodate this irregularity, the model would then fail to reproduce the observed velocity structure of slits 6 to 10.  

The faint emission from beyond the extent of the bright ring is
generally consistent with a bipolar structure, although
the brightness asymmetry between the Northeast and Southwest of the
nebula is not included in the model. Neither the relatively bright, double
velocity component observed at around $-70$\arcsec\ in slits 7 and 8 [Figure \ref{fig:slits6-10}(c) and (d)] nor the 
compact, bright structure appearing on slits 1 and 2 [Figures
  \ref{fig:slits1-5}(a) and (b)] between
$-$70\arcsec{} and $-$90\arcsec{} are reproduced by the model, although
these features are almost certainly associated with the nebula, due
 to their similar velocities.  One possibility is that the compact
 bright feature  marks the nebular rim or opening.  This is, however,
 unlikely, because the faint emission seen at more negative angular
 offsets indicates that the bipolar lobe continues beyond this point.
 It is more likely to be as a result of some filamentary structure in
 the nebular shell.  It is of note that this feature cannot be seen on
 the deep image of SuWt~2 (Figure \ref{fig:slits}), however it is
 unclear whether this is as a result of transmission wavelength
 variations (of the filter) with angle of incidence (and therefore
 field position, see \citealp{ruffle06}) or that it is simply too
 faint. 

 All the PV arrays show what appears to be a very faint component at 0 \kms\ along the full extent of the slit.  This may indicate the presence of a large halo surrounding the nebula, similar to those shown in \cite{corradi03}, or simply be faint galactic emission.

\subsection{Distance estimate and kinematical age}
\label{sec:distance}

There is, as yet, no published distance for SuWt~2, but an estimate of
the distance to the double A-type binary can be made by
  taking $m_v=11.99$ (and assuming both stars are of equal magnitude),
  $E_{b-v}=0.4$ (Exter, priv.\ comm.) and the spectral type as A3V
  ($M_v\sim1.5$, \citealp{astroquant}).
This gives a value of approximately 1 kpc. 

If it is assumed that the nebula is at the same distance as the double A-type binary, then the size of the nebular ring is $\sim 0.44$pc and the lobes even larger. With the same assumption, the expansion velocity of the torus can be converted into a kinematical age of $7600 \pm 1500$ years (or more generally, $7600 \pm 1500$ years kpc$^{-1}$).

\subsection{Systemic velocity of the nebula}
The Heliocentric systemic velocity of the PN was found to be $-25 \pm
5$\kms.  This is somewhat lower than the value of  
$-40 \pm9$ \kms{} found by \citet{west76}.  However, if one considers
that their value is derived from lower spectral-resolution
observations and that the nebula is much brighter on the blue-shifted
side, it is far from unreasonable to conclude that their value was
 taken as the velocity at the brightness peak of the nebula (using this method for our own slits 3 and 8 we determine velocities of $-38 \pm 3$ and $-37 \pm 4$ \kms{}).  Our preferred value of $-25 \pm 5$\kms, however, is determined through comparison of model
spectra to observations (and as such is not biased by the brightness
variations across the nebula).

\section{Discussion and conclusions}

High spatial and spectral resolution longslit \NII\ spectra have been
 obtained from the ring-like PN SuWt~2.  These spectra, together with
 a deep \NII\ image, have been used to derive a spatio-kinematic model
 comprising a bright torus encircling the waist of a fainter bipolar
 nebular shell.  The symmetry axis of the model nebula is inclined at
 $\sim 68\degr$ to the line-of-sight.  A Hubble flow is assumed with
 an expansion velocity for the torus of 28\kms{}, which is consistent
 with typical PNe expansion velocities, further confirming the nature
 of SuWt~2.  The heliocentric systemic velocity of the PN was found to be $-25 \pm
 5$\kms.

The bright double A-type binary is slightly offset from the geometric centre 
of the nebula (\citealp{smith07} and this paper), which may indicate that it is not actually the central 
star, although it is not uncommon to find CSPN offset from their
nebular centres, particularly in ring nebulae like SuWt~2 \citep{pereira08}.  The eclipsing nature of the binary 
\citep{exter03b} implies that the inclination of the binary plane 
is $\sim90\degr$, which is substantially different from the 
inclination angle of the nebula.  
This would seem to rule out 
either star as the nebular progenitor, if it is believed that the
orientation of the nebula follows that of the central binary system   
(e.g.\ \citealp{mitchell07b}).  Moreover, the binary is a double 
A-type system meaning neither component can be the nebular progenitor 
without invoking a born-again scenario (considered unlikely due to 
the determined parameters of the stars, \citealp{bond02}).  

However, it is possible that the nebula was formed from the third
component of a triple system \citep{bond02}. 
  In this case, the plane in which the A-type 
binary orbits the progenitor CSPN would be expected to be similar 
to the nebular inclination. 
One would also expect to see a periodic variation in 
the systemic velocity of the A-type system (as it orbits the third
star).  Exter (paper in preparation) finds that this systemic
velocity does show some variation (priv.\ comm.), with a maximum observed deviation of $\sim21$\kms\ (corresponding to a systemic velocity of $-4$\kms) from the nebular recession velocity of  
$-25$\kms\ reported in this paper.  This indicates that
an association between the nebula and the AA binary (with possible
progenitor companion) cannot be ruled out 
on purely kinematic grounds.  A triple-progenitor scenario is also 
supported by the unusually slow rotation period seen in the A-type 
binary, which could have resulted from an interaction with a third
star \citep{bond08}. However, as yet there is no direct observational
evidence for such a third component.

It has previously been noted that a 
double A-type system would not provide sufficient ionisation to 
illuminate the nebula.  \citet{smith07} speculate that the bright 
star to the Northeast of the nebula could be the ionising source.
However, if the double A-type binary has a wide 
companion white dwarf, which would have been the nebula progenitor 
\citep{bond02}, this would obviously provide the ionising source 
for the nebula.
Alternatively the CSPN may have faded such that 
the nebula is seen in recombination \citep{exter03b}.  This would 
imply that the nebula is very old.  
It is difficult to estimate the 
age of SuWt~2 as its distance is unknown, however, based on the  
distance of 1~kpc to the double A-type binary derived in Section \ref{sec:distance}, the kinematical age of the ring, assuming it is 
radially expanding at 28\kms , is $\sim 7600$ years which is not 
particularly old for a PN. Therefore if the central star has faded to the extent that it can no longer ionise the surrounding nebula, this would imply that SuWt 2 is considerably older and hence lies much further away.  Although this could offer a possible explanation as to why the central star has yet to be directly observed, it is, however, very unlikely as at such a distance the physical size of the nebula would be extremely large.

A connection  between the double A-type binary and
SuWt~2 cannot be ruled out and a detailed radial velocity study of
the binary system would hopefully shed more light on the nature of the
relationship between the two. 

\section*{Acknowledgments}

We would like to thank Dr.\ Katrina Exter for fruitful discussions about the relative systemic velocities of the nebula and AA system.  We also thank the anonymous referee for useful comments which have improved the content and clarity of this paper.

D.J. gratefully acknowledges the support of STFC through his studentship.  N.M.H.V. gratefully acknowledges financial support from ANR; SiNeRGHy grant number ANR-06-CIS6-009-01.

\bibliographystyle{mn2e}
\bibliography{literature.bib}

\label{lastpage}

\end{document}